# Aharonov-Bohm Superperiod in a Laughlin Quasiparticle Interferometer

F. E. Camino, Wei Zhou, V. J. Goldman

*Department of Physics, Stony Brook University, Stony Brook, NY 11794-3800, USA*

We report an Aharonov-Bohm superperiod of five magnetic flux quanta ($5h/e$) observed in a Laughlin quasiparticle interferometer, where an edge channel of the 1/3 fractional quantum Hall fluid encircles an island of the 2/5 fluid. This result does not violate the gauge invariance argument of the Byers-Yang theorem because the magnetic flux, in addition to affecting the Aharonov-Bohm phase of the encircling 1/3 quasiparticles, creates the 2/5 quasiparticles in the island. The superperiod is accordingly understood as imposed by the anyonic statistical interaction of Laughlin quasiparticles.

The fundamental particles exist in three spatial dimensions, and thus all have either bosonic or fermionic integer statistics $\Theta$. Upon execution of a closed loop, both boson and fermions acquire a phase factor of +1, which is unobservable, and thus is usually omitted in an analysis of an interference experiment, such as the Aharonov-Bohm effect. Interference of particles having fractional statistics [1, 2], anyons, would contribute a nontrivial phase $\exp(i2\pi\Theta)$, which ought to be explicitly included in an analysis. Particles comprising collective excitations of a nontrivial system of many integer statistics particles confined to move in 2D can have such anyonic braiding statistics. In particular, the elementary charged excitations (Laughlin quasiparticles, LQPs) of a fractional quantum Hall (FQH) electron fluid [3, 4] have fractional electric charge [4 - 6] and are expected to obey fractional statistics. [7, 8] It is possible to assign definite statistics to LQPs of certain simple FQH fluids based only on the same assumptions that allow to assign their charge. [9] The LQPs of the main FQH sequence at filling $f = p/(2p+1)$, with $p = 1, 2, 3\ldots$, have charge $q = e/(2p+1)$, and their braiding statistics is expected to be $\Theta = 2/(2p+1)$. Several theoretical studies pointed out that the statistics of LQPs can be observed experimentally in variants of the Aharonov-Bohm (AB) effect, [10, 11] but the direct experimental evidence has been lacking.

Our present experiment utilizes a novel LQP interferometer, where an $e/3$ LQP of the $f = 1/3$ FQH fluid executes a closed path around an island of the $f = 2/5$ fluid, Fig. 1. The interference fringes are observed as conductance oscillations as a function of the magnetic flux through the island, that is, the AB effect. We observe the AB period $\Delta\Phi = 5h/e$, equivalent to excitation of ten $q = e/5$ LQPs of the 2/5 fluid. Such "superperiod" of $\Delta\Phi > h/e$ has never been reported before in any system, and is forbidden by the gauge invariance for a true AB geometry, where magnetic flux is added to a region of electron vacuum, as shown by Byers and Yang. [12] Our results do not violate the gauge invariance argument of the Byers-Yang theorem because the flux, in addition to affecting the Aharonov-Bohm phase of the encircling 1/3 quasiparticles, creates the 2/5 quasiparticles in the island. The AB superperiod is accordingly understood as imposed by the anyonic statistical interaction of Laughlin quasiparticles.

The electron interferometer samples were fabricated from low disorder AlGaAs/GaAs heterojunctions. After a shallow 140 nm wet etching, Au/Ti gate metal was deposited in etch trenches, followed by lift-off, Fig. 1(a, b). Samples, mounted on sapphire substrates with In metal (serves as the backgate), were cooled to 10.2 mK in a dilution refrigerator. Four-terminal resistance $R_{XX} \equiv V_X / I_X$ was measured using a 100 pA, 5.4 Hz *ac* current.

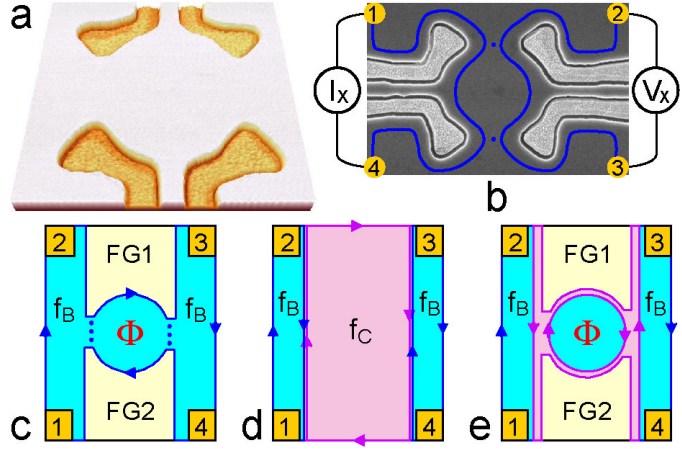

FIG. 1. The interferometer sample. (a),(b) Atomic force and scanning electron micrographs of a typical device. Front gates (FG) are deposited in shallow etch trenches. The depletion potential of the etch trenches defines the central island of 2D electrons (lithographic radius $R \approx 1{,}050$ nm). The backgate (not shown) extends over the entire sample on the opposite side of the insulating GaAs substrate. (c) Schematic of the sample when there is only one QH filling $f$: $f_C$ in the constrictions is equal to $f_B$ in the 2D bulk and in the island. The numbered rectangles are Ohmic contacts. The chiral edge channels follow equipotentials at the periphery of the undepleted 2D electrons; tunneling paths are shown by dots. A closed edge channel path gives rise to AB oscillations in conductance. (d) A QH sample with two fillings exhibits quantized resistance $R_{XX} = (h/e^2)(1/f_C - 1/f_B)$. Observation of a quantized $R_{XX}$ plateau provides definitive values for both $f_C$ and $f_B$. (e) The sample with $f_C < f_B$. There is a quantized $R_{XX}(B)$ plateau, and the sample exhibits AB oscillations as a function of the flux through the *inner* edge ring.

The etch trench depletion potential defines two wide (1,200 nm) constrictions, which separate an approximately circular electron island from the 2D bulk. In this work, the front gate voltages $V_{FG}$ are small, only fine tuning the constrictions for symmetry. The electron density profile $n(r)$ in a circular island defined by etch trenches is evaluated following Ref. 13. [14] For the 2D bulk density $n_B = 1.2 \times 10^{11}$ cm$^{-2}$, there are ~1,700 electrons in the island. The depletion potential has a saddle point in the constrictions, and so has the resulting density profile. In a quantizing magnetic field, the tunneling between the counterpropagating edge channels (possible only over a few magnetic lengths $\ell_0$) occurs near the saddle points. Thus, when AB oscillations are observed, the island edge channel filling is determined by the saddle point filling. From the magnetotransport, the saddle point density in the constrictions $n_C \approx 0.75 \, n_B$.

The local Landau level filling $\nu \equiv hn/eB$ is proportional to $n$, consequently the constriction $\nu_C$ is lower than the bulk $\nu_B$ by ~25% in a given $B$. While $\nu$ is a variable, the QH exact filling $f$, defined via the quantized Hall resistance as $f \equiv h/e^2 R_{XY}$, is a quantum number. Thus there are two regimes possible: one when the whole sample has the same QH filling $f$, and another when there are two QH fillings: $f_C$ in the constrictions, and $f_B$ in the center of the island and in the 2D bulk. For example, there is a range of $B$ where $f_C = f_B = 1$, illustrated in Fig. 1(c). The second regime $f_C < f_B$, Fig. 1(d, e), results in a *quantized* value [15 - 17, 6] of $R_{XX} = (h/e^2)(1/f_C - 1/f_B)$. A quantized plateau in $R_{XX}(B)$ implies QH plateaus for both the constrictions and the bulk, overlapping in a range of $B$ and, in practice, provides definite values for both $f_C$ and $f_B$. Such $R_{XX}(B)$ plateau occurs at $B \approx 12.35$ T for $f_C = 1/3$ and $f_B = 2/5$ ($n_C \approx 0.8 n_B$). However, $f_C = 1$, $f_B = 2$, requiring $n_C \approx 0.5 n_B$, is not possible in this sample.

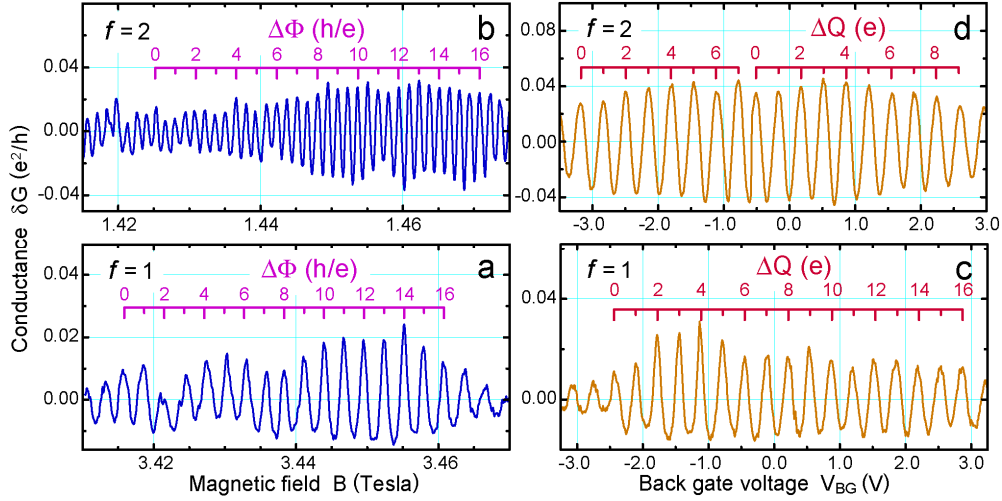

FIG. 2. Interference of electrons in the integer QH regime. (a, b) AB oscillations in conductance when one ($f = 1$) or two ($f = 2$) Landau levels are filled. The flux period $\Delta\Phi = h/e$ gives the outer edge ring radius 685 nm. (c, d) Positive $V_{BG}$ attracts 2D electrons one by one to the area within the AB path, modulating the conductance. This calibrates the increment $\Delta V_{BG}$ needed to increase the charge by $\Delta Q = e$. Note that $\Delta V_{BG}$ is independent of $f$.

*The integer regime.* The relevant particles are electrons of charge $e$ and integer statistics, thus, we obtain an absolute calibration of the ring area and the backgate action of the interferometer. Fig. 2 shows AB oscillations for $f_C = f_B = 1$ and 2. Conductance variation $\delta G = \delta R_{XX}/R_{XY}^2$ is calculated from the $R_{XX}$ data after subtracting a smooth background. The AB ring is formed here by the edge channel circling the island, including two quantum tunneling links, Fig. 1(c). The $f = 1$ period $\Delta B_1 \approx 2.81$ mT gives the area of the *outer* edge ring $S_O = h/e\Delta B_1 \approx 1.47$ µm², the radius $r_O \approx 685$ nm. The $f = 2$ period is very close: $2\Delta B_2 \approx 2.85$ mT gives the area $S_O \approx 1.45$ µm². The $f = 2$ fundamental period contains two oscillations, $2\Delta B_2 S_O = h/e$, because there are two filled spin-polarized Landau levels. We calibrate the backgate action $\delta Q/\delta V_{BG}$, where $Q$ is the charge of electrons within the AB path, because the density in the island is not expected to increase by precisely the same amount as $n_B$, unlike a quantum antidot (the antidot is completely surrounded by a QH fluid). [6, 17] Fig. 2(c, d) shows oscillations as a function of $V_{BG}$ for $f_C = f_B = 1$ and 2. The period $\Delta V_{BG}$ corresponds to change $\Delta N = 1$ in the number of electrons within the AB path. Thus $\Delta V_{BG}$ should be the same for all spin-polarized IQH states, provided the radius of the edge ring is constant; indeed, $\Delta V_{BG} = 332$ mV for $f = 1$ and 342 mV for $f = 2$.

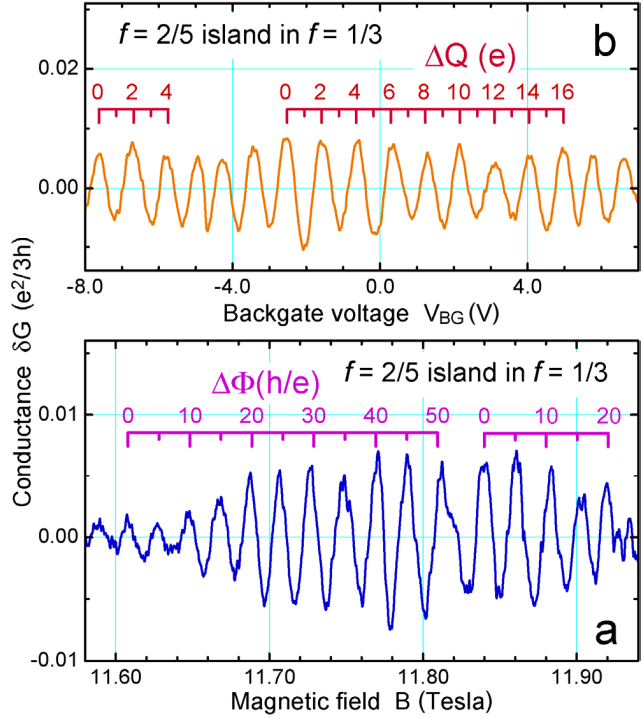

FIG. 3. Aharonov-Bohm interference of $e/3$ LQPs circling an island of $f = 2/5$ FQH fluid. (a) Flux through the island period $\Delta \Phi = 5h/e$ corresponds to creation of ten $e/5$ LQPs in the island [one $h/e$ excites two $e/5$ LQPs from the $f = 2/5$ FQH condensate, the total (LQPs + condensate) charge is fixed]. Such "superperiod" $\Delta \Phi > h/e$ has never been reported before. (b) The backgate voltage period of $\Delta Q = 2e$ directly confirms that the $e/3$ LQP consecutive orbits around the 2/5 island are quantized by a condition requiring increment of ten $e/5$ LQPs.

*The FQH regime*. We focus on the regime when an 1/3 annulus surrounds an island of the 2/5 FQH fluid, Fig. 1(e). Here, we observe AB oscillations with period $\Delta B \approx 20.1$ mT, Fig. 3(a). The period gives the *inner* edge ring area $S_I = 5h/e\Delta B \approx 1.03$ µm$^2$, radius $r_I \approx 570$ nm. Fig. 3(b) shows the oscillations as a function of $V_{BG}$, the period $\Delta V_{BG} \approx 937$ mV. We are confident that current flows through $f_C = 1/3$ region separating two $f_B = 2/5$ 2D regions with Ohmic contacts because $R_{XX}(B)$ exhibits a plateau at $\frac{1}{2}h/e^2$, Fig. 4(a). The island center density is 4% less than $n_B$; [14] thus island $\nu$ the same as $\nu_B$ occurs at 4% lower $B$. The ratio of the periods $\Delta B / \Delta V_{BG} \propto N_\Phi / N_e = 1/f$ is independent of the AB ring area. $N_\Phi$ and $N_e$ are the number of flux quanta and electrons within the AB path area. The fact that the ratios fall on a straight line forced through zero confirms the island filling $f = 2/5$ at 11.9 T, Fig. 4(b).

The striking feature of the oscillations in Fig. 3(a) is the AB period of five fundamental flux quanta: $\Delta\Phi = 5h/e$! To the best of our knowledge, such superperiod of $\Delta\Phi > h/e$ has never been reported before. Addition of flux $h/e$ to an area occupied by the 1/3 FQH condensate creates a vortex, an $e/3$ quasihole [4]. Likewise, addition of flux $h/e$ to the 2/5 FQH fluid creates two vortices, that is, two $e/5$ quasiholes [5]. These predictions have been verified at a microscopic level in quantum antidot experiments [6, 17]. Thus, addition of $5h/e$ to the $f = 2/5$ island creates ten $e/5$ LQPs with total charge $\Delta Q = 2e$, confirmed by the backgate data, Fig. 3(b). In contrast, the periods observed in quantum antidots correspond to addition of one LQP only, both for the 1/3 and 2/5 cases. The principal difference between the present interferometer and the antidots is that in quantum antidots the FQH fluid surrounds electron vacuum, while in the present interferometer the 1/3 fluid surrounds an island of the 2/5 fluid. The gauge invariance argument [12] requiring $\Delta\Phi \leq h/e$ for the true AB geometry is not applicable here because the interior of the AB path contains electrons, and applied flux creates LQPs in the island. Addition of flux does excite LQPs, the system is not the same as prior to flux addition, thus the applied flux can not be annulled by a singular gauge transformation. Likewise, Laughlin's "gedanken experiment" can not be applied to the interferometer geometry to assume that any charge is transferred in or out of the island by the AB flux.

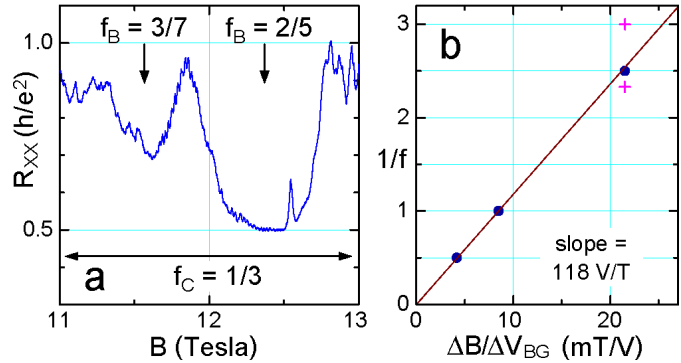

FIG. 4. (a) Magnetoresistance of the interferometer sample at 10.2 mK. The horizontal arrow show approximately the $f_C = 1/3$ plateau. Note the *quantized* plateau $R_{XX}(B) = h/2e^2$ at 12.35 T, obtainable only with $f_C = 1/3$, $f_B = 2/5$. (b) The oscillation period ratio for the data of Figs. 2 and 3. $\Delta B / \Delta V_{BG} \propto 1/f$, independent of the AB path area. The straight line goes through (0,0) and the $f = 1$ point. Experimental $\Delta B / \Delta V_{BG} = 21.4$ mT/V gives the island filling $f = 2/5$. The crosses (the nearest FQHE $f = 3/7$ and 1/3) do not fit the data.

A*lternative interpretations*. Any viable interpretation must conform to the experimental facts: (i) current is transported by $e/3$ LQPs of the surrounding 1/3 fluid, Fig. 4(a). (ii) The oscillations originate in the 2/5 island, Fig. 4(b). (iii) Both flux (no systematic net charging of the island) and backgate (systematic charge transfer into the island) periodic oscillations, must be accounted for at least 10 periods away from exact filling, Fig. 3. (iv) The oscillations are found to be robust, observed in four distinct cooldowns, persisting to 140 mK and upon application of a moderate front gate voltage ±300 mV [18]; (v) We have observed similar integer and fractional AB data in another sample with a larger lithographic $R \approx 1,300$ nm [14]. $\Delta B_{2/5} \approx 6.4 \Delta B_1$ is consistent, upon the same depletion potential analysis, with the period $\Delta\Phi = 5h/e$.

A different $r_I$ would yield a different $\Delta\Phi$. We restrict analysis to $\Delta\Phi$ of simple rational multiples of $h/e$, consistent with the particle nature of charged elementary excitations. The observed field period $\Delta B_{2/5} \approx 7.15 \Delta B_1$, the facts that current flows in the outer edge channels with $f_C = 1/3$, and that experimental $\Delta B / \Delta V_{BG}$ gives the filling 2/5, are all consistent with formation of an $f = 2/5$ island within the 1/3 outer edge ring, the inner ring radius $r_I < r_O$. This is also expected since the island center density $n_I \approx 1.22 n_C$, just above the assigned $f_I \approx 1.20 f_C$. The alternative $\Delta\Phi = \frac{5}{2} h/e$, $\Delta Q = e$ (still constituting an AB superperiod) is ruled out as giving too small $r_I \approx 400$ nm. The confining potential at $r_I \approx 400$ nm is simply too weak to support a *stable* edge ring; an estimate using models [13,14] gives an order of magnitude weaker gradient $dn(r)/dr$ at 400 nm than at 570 nm. The alternative of $\Delta\Phi = h/e$, $\Delta Q = \frac{2}{5} e$ gives yet smaller $r_I = 255$ nm, where confining potential is nearly flat and thus cannot define an edge channel. Another consideration is tunneling through the distance $t = r_O - r_I$, between the inner and the outer edge channels. The LQP tunneling rate is estimated [19] as $\sim \exp[-(t/2\sqrt{3\pi}\ell_0)^2]$, which gives $10^{-2}$ for $r_I \approx 570$ nm (agreeing with experiment, Fig. 3) and $10^{-14}$ for $r_I \approx 400$ nm, much too small to observe.

Exchange of charge between the island and the surrounding FQH fluid in increments of one LQP, $\Delta Q = e/5$ (as in quantum antidots) is clearly not consistent with the data. A model where no LQPs are created (only exact filling FQH condensates are considered), but instead the 1/3-2/5 condensate boundary shifts [20], is not energetically feasible. As is well known, in a large 2D FQH fluid, changing $\nu$ away from the exact filling $f$ is accompanied by creation of LQPs, so as to maintain average charge neutrality; the ground state consists of an $\nu = f$ condensate and the matching density of LQPs [4,5,7,17]. Forcing exact filling (and no LQPs) at $B \neq B_f \equiv hn/ef$ changes $n$ away from the equilibrium value determined by the positively charged donors. In present geometry, this would lead to formation of a charged $\nu = f = 2/5$ disc surrounded by an oppositely charged 1/3 annulus, and thus huge Coulomb energy. For the tenth oscillation from the exact filling, the net charge is $20e$, the charging energy ~1,000 K, much more than the LQP gap. An intermediate model where LQPs are allowed, but are envisioned concentrated near the 1/3-2/5 boundary, besides still present charging energy (additional to the equilibrium ground state energy), must overcome the difficulty of the local $\nu$ being affected. Concentrating 100 $e/5$ LQPs within $5\ell_0$ of $r_I$ changes local filling to $\nu = 0.528 > \frac{1}{2}$, well outside the 2/5 plateau. This would certainly break the observed AB periodicity. Note that Eq. (1), being of a topological nature, is not sensitive to the precise position of LQPs, so long as the island remains on the $f = 2/5$ plateau.

*The Berry phase* $\gamma$ of an $e/3$ LQP encircling a closed path in the 1/3 FQH condensate was calculated in [8]. The difference between an "empty" and a loop containing another LQP is $2\pi\Theta_{1/3} = 4\pi/3$, identified as the statistical contribution. It is instructive to consider a quantum antidot. When the chemical potential moves between two successive LQP states, the Berry phase period $\Delta\gamma$ is $2\pi$: $\Delta\gamma = \frac{q}{\hbar}\Delta\Phi + 2\pi\Theta\Delta N = 2\pi$ [10, 21]. When occupation of the antidot changes by one $e/3$ quasihole [22], $\Delta N = 1$, the experiments give $\Delta\Phi = h/e$. Thus $\Delta\gamma = 2\pi(q/e + \Theta_{1/3}) = 2\pi$ only if LQPs have anyonic $\Theta_{1/3} = 2/3$. This is not entirely satisfactory as a *direct* demonstration of the fractional statistics, because in a quantum antidot the tunneling LQP encircles electron vacuum, and the most important ingredient, the fact that the period $\Delta\Phi = h/e$, not $h/q$, is ensured by the Byers-Yang theorem.

In the interferometer, if we neglect the symmetry properties of the FQH fluids, in the absence of a Coulomb blockade, there is no *a priori* constraint that the total charge of the 2/5 island be quantized in units of $e$, much less in units of $2e$. The island fluid could adjust in increments of one LQP, any fractional charge imbalance supplied from the contacts. Thus the periods $\Delta\Phi = 5h/e$, $\Delta Q = 2e$ must be imposed by the symmetry properties of the two FQH fluids. The current used to measure conductance is transported by LQPs of the outside 1/3 fluid; therefore, we infer that $\Delta\Phi = 5h/e$ results from the $2\pi$ periodicity of the Berry phase of the $q = -e/3$ quasielectron encircling $\Delta N = 10$ of $e/5$ quasiholes of the $f = 2/5$ fluid:

$$\Delta\gamma = \frac{q}{\hbar}\Delta\Phi + 2\pi\,\Theta^{1/3}_{2/5}\Delta N = 2\pi\ . \tag{1}$$

Solving Eq. (1) gives the relative statistics $\Theta^{-1/3}_{2/5} = 4/15$, and using the quasihole $q = e/3$ gives $\Theta^{1/3}_{2/5} = -1/15$. We are not aware of a theoretical work explicitly evaluating $\Theta^{1/3}_{2/5}$, but the value is consistent with the flux attachment models [2].

In conclusion, we realized a novel Laughlin quasiparticle interferometer, where an $e/3$ LQP executes a closed path around an island of the 2/5 FQH fluid. The central results obtained, the Aharonov-Bohm superperiods of $\Delta\Phi = 5h/e$ and $\Delta Q = 2e$ are robust. These results do not violate the Byers-Yang theorem, and are interpreted as implying anyonic braiding statistics of LQPs.

We thank D. V. Averin for discussions. This work was supported in part by the NSF, and by US NSA and ARDA through US ARO.